\documentclass[prd,preprint,nofootinbib]{revtex4-1}

\usepackage{amssymb,amsmath}
\usepackage{color} 
\usepackage{graphicx} 
\usepackage{float} 
\usepackage{braket}

\usepackage{xspace}

\newcommand{\mH}{\text{H}}
\newcommand{\mHbar}{\bar{\text{H}}}
\newcommand{\mHplus}{\text{H}^{\text{+}}}
\newcommand{\mHminus}{\text{H}^{\text{-}}}

\newcommand{\meminus}{\text{e}^{\text{-}}}
\newcommand{\mHtwo}{\text{H}_2}

\newcommand{\Hbar}{$\bar{\text{H}}$\xspace}
\newcommand{\Hplus}{$\text{H}^{\text{+}}$\xspace}
\newcommand{\Hminus}{$\text{H}^{\text{-}}$\xspace}

\newcommand{\eminus}{$\text{e}^{\text{-}}$\xspace}
\newcommand{\Htwo}{$\text{H}_2$\xspace}

\newcommand{\HI}{H\textsc{i}\xspace}
\newcommand{\HII}{H\textsc{ii}\xspace}

\newcommand\unuse[1]{}

\def\beq{\begin{equation}}
\def\eeq{\end{equation}}
\def\beqa{\begin{eqnarray}}
\def\eeqa{\end{eqnarray}}
\def\ben{\begin{enumerate}}
\def\een{\end{enumerate}}

\usepackage[T1]{fontenc} 

\begin{document}

\title{Re-visiting the bounds on hydrogen-antihydrogen oscillations from diffuse $\gamma$-ray surveys}
\author{Yuval Grossman}
\email{yg73@cornell.edu}
\affiliation{Laboratory for Elementary-Particle Physics, Cornell University, Ithaca, N.Y.}
\author{Wee Hao Ng}
\email{wn68@cornell.edu}
\affiliation{Laboratory for Elementary-Particle Physics, Cornell University, Ithaca, N.Y.}
\author{Shamayita Ray}
\email{shamayita.ray@gmail.com}
\affiliation{Department of Physics, University of Calcutta, Kolkata 700 009, India}

\begin{abstract}
Surveys of diffuse $\gamma$-ray in the interstellar medium (ISM) can be used to probe hydrogen-antihydrogen oscillations, by detecting the $\gamma$-ray emission from antihydrogen annihilation. A bound on the oscillation parameter $\delta$ was originally derived by Feinberg, Goldhaber and Steigman (1978). In this paper, we re-visit the original derivation by performing a more detailed analysis that (1) incorporates suppression effects from additional elastic and inelastic processes, (2) treats the ISM as a multi-phase medium, and (3) utilises more recent $\gamma$-ray data from the \textit{Fermi} Large Area Telescope. We find that suppression from elastic scattering plays a more important role than previously thought, while the multi-phase nature of the ISM affects how the $\gamma$-ray data should be utilised. We derive a more accurate bound on the oscillation period that is about an order of magnitude weaker than the older bound.
\end{abstract}

\maketitle

\section{Introduction}

At the classical level, baryon ($B$) and lepton ($L$) numbers are conserved quantities in the Standard Model (SM). One of Sakharov's condition \cite{Sakharov:1967dj} for a dynamical explanation of the baryon asymmetry in the universe requires that $B$ conservation be violated. Mechanisms like electroweak baryogenesis \cite{Kuzmin:1985mm} or leptogenesis \cite{Fukugita:1986hr} achieve this through sphaleron processes that makes use of $B+L$ violation in the SM at the quantum level, while mechanisms like baryogenesis in the Grand Unified Theories (GUT) \cite{Nanopoulos:1979gx} introduce processes that directly violate $B$ at the classical level. However, proton decay imposes strong constraints on models that directly allow $\Delta B = \Delta L = 1$ processes. One intriguing possibility is to consider models \cite{Mohapatra:1982aj, FileviezPerez:2011dg} where proton decay is forbidden/suppressed, but yet allow processes with $\Delta B = 2$ or $\Delta B = \Delta L = 2$ to occur. In these cases, processes such as neutron-antineutron oscillations \cite{Phillips:2014fgb}, $pp \to e^+e^+$ annihilations \cite{Bramante:2014uda} or hydrogen-antihydrogen (H-\Hbar) oscillation may become more important probes of $B$ violation. In this paper we concentrate on H-\Hbar oscillation.

One way to detect H-\Hbar oscillations is through $\gamma$-rays from the annihilation of \Hbar with other particles in its vicinity (henceforth called ``oscillation-induced $\gamma$-rays''). A good place to look for this is the interstellar medium (ISM), first because of the immense amount of atomic hydrogen present, and second because the low density allow a larger oscillation amplitude and hence a larger proportion of \Hbar to exist than in terrestrial sources. These $\gamma$-rays then show up in diffuse $\gamma$-ray surveys on top of other $\gamma$-ray emitting processes, such as cosmic ray (CR) interaction with matter. This idea is not new and a bound on the oscillation was first derived in \cite{Feinberg:1978sd}. The goal of the present paper is to revisit the bounds, for the following reasons. 
\begin{enumerate}
\item
In the original derivation, the amplitude of oscillation was assumed to be limited by H-\Hbar annihilation. However, we do not know \textit{a priori} how this compares to the effects of other processes such as elastic scattering.

\item
We now have a better understanding of the phases of the ISM, $\gamma$-ray production within the ISM, as well as updated $\gamma$-ray survey results from the \textit{Fermi} Large Area Telescope (LAT).

\item
Finally, many steps are involved in deriving the experimental bounds on the oscillations. While we are only interested in an order-of-magnitude estimate, we want to reduce the uncertainty in each step as much as possible to avoid having the cumulative errors become too large. Therefore, besides improving on the oscillation and ISM model, we also want to utilise updated parameter values from literature rather than just rely on crude estimates.
\end{enumerate}

This paper is structured as follows. In Sec.~\ref{sec:formalism}, we present a model that describes H-\Hbar oscillations in a medium, and use the model to derive a formula for the oscillation-induced $\gamma$-ray emissivity. In Sec.~\ref{sec:ism-and-parameter}, we use this formula, together with available data for various elastic and inelastic processes, to calculate the emissivities of the relevant phases of the ISM. It then allows us in Sec.~\ref{sec:bounds} to obtain a bound on the oscillation parameter $\delta$ based on the \textit{Fermi} LAT data presented in Ref.~\cite{Abdo:2009ka}. We conclude in Sec.~\ref{sec:discussion} with a comparison of our bound with that from other $\Delta B=\Delta L =\pm 2$ processes. To keep the text focused, most technical details have been placed in the appendices.

\section{Model of H-\Hbar oscillation} \label{sec:formalism}

To infer the oscillation-induced $\gamma$-ray emissivity, we need to know the probability of an H atom in the ISM becoming a \Hbar. This in turn can be derived from a single-atom model of H-\Hbar oscillation. The vacuum formalism is very straightforward; however the main issue here is to account for interactions with the environment. Some of the effects are well-understood: for example, forward scattering gives rise to coherent matter effects known from neutrino oscillations, while inelastic processes such as \Hbar annihilation cause the state to leave the Hilbert space of interest and hence their effects are analogous to decays in meson oscillations. Both of these effects can be taken care of by modifications to the effective Hamiltonian.

Less well-recognised are effects that require going beyond the effective Hamiltonian, and require a density matrix formalism. First, say H and \Hbar have different elastic scattering amplitudes off the same target, i.e.~$f(\theta) \ne \bar{f}(\theta)$, where $\theta$ is the angle of scattering. Then non-forward scattering cause the identity of the atom (H or \Hbar) to become entangled with its momentum and hence a two-level pure state formalism does not work if we want to incorporate elastic scattering beyond just forward scattering. Also, since the scattering environment is usually random, even a pure state formalism incorporating both identity and momentum degrees of freedom is insufficient. Second, chemical reactions such as recombination generate new ``unoscillated'' H atoms to replenish those lost to inelastic processes. Since these reactions should be treated as classical source terms, again a density matrix formalism is required. The model we adopt is similar to the original Feinberg-Weinberg model \cite{Feinberg:1961zza} that was also used in \cite{Feinberg:1978sd}. We then extend it to take into account more general sources of suppression. We also highlight the differences between our work and that of \cite{Feinberg:1978sd}.

\subsection{Model description} \label{sec:model-description}

We regard H and \Hbar as basis states of a two-level system (Hilbert space $\mathcal{H}_A$). In principle, there are other degrees of freedom such as momentum, atomic level and spin (Hilbert space $\mathcal{H}_B$), but since we are only interested in finding the probability of being \Hbar, we trace them out in the full density matrix $\rho_{\text{full}}(t)$ to obtain a reduced $2 \times 2$ density matrix $\rho(t)$. The quantum kinetic equation of $\rho(t)$ will then depend on the moments of the other degrees of freedom, e.g.~$\text{Tr}_B[p^2\rho_{\text{full}}(t)]$, and is hence not closed. To close this equation, we replace, say, the example above by $\langle p^2(t)\rangle \rho(t)$, and assume that $\langle p^2(t)\rangle$ is just given by the present-day value (since we are only interested in a quasi-steady solution). Also, since most of the atoms in the ISM phases of interest are in the $1S$ state, any average involving atomic level and spin is equivalent to a $1S$ hyperfine average.

\subsubsection{Elastic scattering}
First, we take into account elastic scattering of the atom with other particles (targets). Let $i$ denote the target species. Then $\rho(t)$ satisfies the kinetic equation \cite{Feinberg:1961zza}
\begin{equation}
\partial_t \rho(t) = -i [H \rho(t) - \rho(t)H^\dagger] + \sum_i\left[ n_i v_i \int d\Omega F_i(\theta) \rho F_i^\dagger(\theta)\right],
\label{eq:formalism-elastic}
\end{equation}
where
\begin{equation}
H \equiv \begin{pmatrix}
E - \sum_i\left[\frac{2\pi n_i v_i}{p_i} f_{i,p_i}(0)\right] & \frac{\delta}{2} \\
\frac{\delta^*}{2} & E - \sum_i\left[\frac{2\pi n_i v_i}{p_i} \bar{f}_{i,p_i}(0)\right]
\end{pmatrix}, \quad
F_i(\theta) \equiv \begin{pmatrix} f_{i,p_i}(\theta) & 0 \\ 0 & \bar{f}_{i,p_i}(\theta) \end{pmatrix},
\end{equation}
and the symbols used here are defined as follows:
\begin{itemize}
\item $E$: the mean energy of an atom in vacuum (equal for H and \Hbar by CPT) in the ISM rest frame,
\item $n_i$: the number density of species $i$,
\item $v_i$: the r.m.s.~speed of approach between atom and a species $i$ particle,
\item $p_i$: the r.m.s.~momentum in centre-of-mass frame of the atom and a species $i$ particle,
\item $f_{i,p_i}(\theta)$ ($\bar{f}_{i,p_i}(\theta)$): scattering amplitude of H (\Hbar) off a species $i$ particle with momentum $p_i$ in centre-of-mass frame, and 
\item $\frac{\delta}{2}$: off-diagonal matrix element generated by $\Delta B=\Delta L = \pm 2$ operators.
\end{itemize}

The assumptions involved are presented in App.~\ref{app:formalism}. We just explain a few features of Eq.~(\ref{eq:formalism-elastic}) here. The first term describes the usual time-evolution with an effective non-Hermitian Hamiltonian $H$, comprising the energy $E$ of the atom in vacuum, the oscillation term $\delta$, and coherent forward scattering $f_{i,p_i}(0)$ and $\bar{f}_{i,p_i}(0)$, summed over all target species $i$. Differences in $f_{i,p_i}(0)$ and $\bar{f}_{i,p_i}(0)$ can suppress the oscillations, just like coherent matter effects in neutrino oscillations. The 
optical theorem ensures that even for elastic scattering $f_{i,p_i}(0)$ and $\bar{f}_{i,p_i}(0)$ are complex quantities, with the imaginary parts related to the total scattering rate. As a result, time evolution under the first term alone cause the total probability represented by $\text{Tr}(\rho)$ to decrease. This decrease is analogous to the effects of the ``out'' collision term in Boltzmann transport equation. Probability conservation is restored by the second term, analogous to the ``in'' collision term.

\subsubsection{Inelastic and production processes} \label{sec:inelastic}

To complete the picture, we want to include inelastic processes as well. We argue in App.~\ref{app:inelastic} that among all the inelastic processes, only those where the H/\Hbar atom ``disappears'' are potentially important. This includes ionisation, chemical reactions as well as \Hbar annihilation. Since these processes take the state out of the Hilbert space $\mathcal{H}_A$, they can be represented by imaginary contributions $i\omega_I/2$ and $i\bar{\omega}_I/2$ to the diagonal elements of $H$, where $\omega_I$ ($\bar{\omega}_I$) denotes the total rate of these processes per H (\Hbar) atom.

However, just as H/\Hbar atoms can ``disappear'', they can also ``reappear'' through production process such as recombination and \Htwo dissociation. These processes correspond to source terms for the $\rho_{11}$ matrix element, which we introduce as $\omega_P \rho_{11}$ in Eq.~(\ref{eq_inelastic}). $\omega_P$ can be interpreted as the rate of H production per unit volume, normalised by the number density of H. 
Furthermore, if we assume that the ISM is in a quasi-steady state (approximate ionisation balance, chemical equilibrium, etc.), then this source term can be approximated as $\omega_P \simeq \omega_I$ up to a small difference of order the quasi-steady rate of change.
In principle, we can also include a source term for $\rho_{22}$, e.g.~from re-combination of CR positrons and antiprotons to form \Hbar. However, based on measurements of the CR antiproton flux \cite{Aguilar:2016kjl}, this contribution is expected to be negligible compared to \Hbar production from oscillations at the upper bound of $|\delta|$.

The time-evolution equation is then given by
\begin{equation}
\partial_t \rho = -i [H\rho - \rho H^\dagger] + \sum_i \left[ n_i v_i \int d\Omega F_i(\theta) \rho F_i^\dagger(\theta) \right] + 
\begin{pmatrix}
\omega_P \rho_{11} & 0 \\
0 & 0 \end{pmatrix}
\end{equation}
with a modified effective Hamiltonian
\begin{equation}
H \equiv \begin{pmatrix}
E - \sum_i\left[\frac{2\pi n_i v_i}{p_i} f_{i,p_i}(0)\right] - \frac{i}{2}\omega_I & \frac{\delta}{2} \\
\frac{\delta^*}{2}
& E - \sum_i\left[\frac{2\pi n_i v_i}{p_i} \bar{f}_{i,p_i}(0)\right] - \frac{i}{2} \bar{\omega}_I 
\end{pmatrix}.
\label{eq_inelastic}
\end{equation}

\subsubsection{Reformulating the model}

It is instructive to rewrite $\rho(t)$ as a column vector $\rho(t) \equiv (\rho_{11}, \rho_{12}, \rho_{21}, \rho_{22})^T$ \cite{Feinberg:1978sd}. The time evolution equation then becomes
\begin{equation}
\partial_t \rho(t) = M \rho,
\label{eq:formalism-column}
\end{equation}
where
\begin{equation}
M \equiv \begin{pmatrix}  \omega_P -\omega_I & i\frac{\delta^*}{2} & -i\frac{\delta}{2} & 0 \\
i\frac{\delta}{2} & \epsilon' & 0 & -i\frac{\delta}{2}\\
-i\frac{\delta^*}{2} & 0 & \epsilon^{\prime *}
& i\frac{\delta^*}{2} \\
0 & -i\frac{\delta^*}{2} & i\frac{\delta}{2} &  - \bar{\omega}_I 
\end{pmatrix},
\end{equation}
\begin{equation}
\epsilon' \equiv  i \sum_i n_i v_i \left[ \Delta_i + \int d\Omega \text{Im}(\bar{f}_{i,p_i}^*f_{i,p_i}) \right] - \left[\frac{\omega_I + \bar{\omega}_I}{2} + \sum_i \frac{n_i v_i}{2} \int d\Omega |f_{i,p_i} - \bar{f}_{i,p_i}|^2 \right],
\label{eq:epsilonprime}
\end{equation}
\begin{equation}
\Delta_i \equiv \frac{2\pi}{p_i} \text{Re}[f_{i,p_i}(0)-\bar{f}_{i,p_i}(0)].
\end{equation}

Some observations:
\begin{itemize}
\item If $f_{i,p_i} = \bar{f}_{i,p_i}$, then all instances of $f_{i,p_i}$ and $\bar{f}_{i,p_i}$ vanish from $M$. In other words, elastic scattering does not suppress oscillations unless it can differentiate between H and \Hbar amplitude-wise. This means, for example, that we can ignore elastic scattering with photons.

\item If $\omega_I = \bar{\omega}_I$, then their combined contributions to $M$ is just proportional to the identity, so they only lead to an overall decay factor. Therefore, inelastic processes also do not suppress oscillations unless they can differentiate between H and \Hbar rate-wise.

\item Oscillations are also suppressed by the source term $\omega_P \rho_{11}$, although the physical mechanism is somewhat indirect. Here new H atoms that have yet to oscillate are being added to the system. This suppression is why despite our previous comment, we still need to consider inelastic processes such as photo-ionisation that have the same rate for H and \Hbar, since $\omega_I$ informs us about $\omega_P$ in the quasi-steady state.

\end{itemize}

Note that our formalism here is similar to the one used in \cite{Feinberg:1978sd} (see Eq.~(2.4) there). However, they did not include a source term $\omega_P$, and they also assumed that the only important process is H-\Hbar annihilation. As a result, they have $\bar{\omega}_I \gg \omega_I$ (since it is much easier for a \Hbar to find a H to annihilate with, than vice versa) and $|\epsilon'| \simeq \bar{\omega}_I/2$. In contrast, we do not make the same assumptions but instead consider a wide range of elastic and inelastic processes. 

\subsection{Formula for $\gamma$-ray emissivity}

We want to use our model to derive a formula for the $\gamma$-ray emissivity. To do so, we need to find the solution to Eq.~(\ref{eq:formalism-column}) that best describes a H/\Hbar atom in the ISM, from which we can then obtain the \Hbar number density and hence the emissivity.

Most of the parameters in $M$ depend on the number densities of atomic hydrogen and other species in the ISM, so Eq.~(\ref{eq:formalism-column}) is actually much harder to solve than it seems. However, since we are only interested in the quasi-steady solution, it is actually self-consistent to assume these parameters as constants, at least for timescales short compared to the quasi-steady rate of change. Even though the quasi-steady solution based on this assumption may become inaccurate at longer times, it does not matter since we are using \emph{present-day} parameter values. In other words, the reference starting time is actually the present, so we read off the present-day \Hbar probability $\rho_{22}$ from the solution at $t=0$. 

With this assumption, among the four eigenvectors of $M$, three have eigenvalues with negative real parts of order $|\epsilon'|$ or $\bar{\omega}_I$, while the fourth is given by
\begin{equation}
\lambda = \omega_P - \omega_I + \mathcal{O}(\epsilon^2|\epsilon'|, \epsilon^2|\bar{\omega}_I|)
\end{equation}
where $\epsilon \equiv \text{Max}\left\lbrace \left\vert \frac{\delta}{\epsilon'}\right\vert, \left\vert \frac{\delta}{\bar{\omega}_I}\right\vert \right\rbrace$ is a small parameter. The first three solutions correspond to transients that decay rapidly (although the actual decay rate may be somewhat different since these solutions are not consistent with the assumption about the parameters being constant), while the fourth solution does indeed change at the quasi-steady rate $|\omega_P-\omega_I|$ and is thus the one we want. The corresponding eigenvector is given by
\begin{equation}
v = \begin{pmatrix}
1 + \mathcal{O}(\epsilon^2)\\
-\frac{i\delta}{2(\epsilon'+\omega_I-\omega_P)} + \mathcal{O}(\epsilon^3)\\
\left[-\frac{i\delta}{2(\epsilon'+\omega_I-\omega_P)} + \mathcal{O}(\epsilon^3)\right]^*\\
-\left\vert \frac{\delta}{\epsilon' + \omega_I - \omega_P} \right\vert^2 \frac{\text{Re}(\epsilon' + \omega_I - \omega_P)}{2(\bar{\omega}_I - \omega_I + \omega_P)} + \mathcal{O}(\epsilon^4)
\end{pmatrix}.
\end{equation}
We observe that of the four components, $v_1 \simeq 1$, $v_2 = v_3^* \sim \mathcal{O}(\epsilon)$, and $v_4 \sim \mathcal{O}(\epsilon^2)$.

Since $\tfrac{v_1}{v_1+v_4}$ and $\tfrac{v_4}{v_1 + v_4}$ correspond to the probability of being H and \Hbar, we can estimate the rate of \Hbar annihilation per unit volume as
\begin{equation}
\frac{v_4}{v_1} n_{\mH}n_i \langle \sigma_i v_i\rangle
\simeq -\left\vert \frac{\delta}{\epsilon'} \right\vert^2 \frac{\text{Re}(\epsilon')}{2\bar{\omega}_I}
\bar{\omega}_{\text{ann}}
\end{equation}
where $\bar{\omega}_{\text{ann}}$ is the annihilation rate per \Hbar (we allow it to differ from $\bar{\omega}_I$ in case there are other more important \Hbar ``disappearance'' processes), and we have dropped the much smaller quasi-steady rate $|\omega_P - \omega_I|$ relative to $\bar{\omega}_I$ and $\epsilon'$. This is a positive quantity since $\text{Re}(\epsilon') < 0$.  Note that $\omega_P$ has disappeared completely (it is not present in $\epsilon'$) since its main role is to cancel $\omega_I$ at certain places to give a much smaller quasi-steady rate that can then be neglected.

For comparison with $\gamma$-ray data later, it is useful to convert the previous rate per unit volume into an oscillation-induced emissivity per H atom, which gives
\begin{equation}
\epsilon_{\gamma} = -\frac{g_{\gamma}}{4\pi}\left\vert \frac{\delta}{\epsilon'} \right\vert^2 \frac{\text{Re}(\epsilon')}{2\bar{\omega}_I}
\bar{\omega}_{\text{ann}} \text{ photons } \text{sr}^{-1},
\label{eq:emissivity}
\end{equation}
where $g_{\gamma}$ is the average number of $\gamma$-ray photons emitted in the annihilation. We discuss its value below for specific situations.


\section{Calculating the emissivities} \label{sec:ism-and-parameter}

In the previous section, we derived a formula for the oscillation-induced $\gamma$-ray emissivity per H atom, Eq.~(\ref{eq:emissivity}). To make further progress, we need numerical values of the parameters in this formula, except for the unknown $|\delta|$ that we want to constrain. We begin this section by identifying phases of the ISM that are expected to be the dominant sources of these $\gamma$-rays. Using available data for a wide variety of elastic and inelastic processes, we then calculate the parameter values and hence the emissivity for each phase. We adopt the standard astronomical notation of \HI and \HII for atomic and ionised hydrogen.

\subsection{Phases of the ISM} \label{sec:ism}

The \textit{Fermi} LAT data presented in Abdo \textit{et al.} \cite{Abdo:2009ka} focuses on $\gamma$-ray emission from \HI and is hence of particular relevance to our work. We want to consider the same sector of the ISM, bounded by Galactic longitude $200^\circ < l < 260^\circ$, and latitude $22^\circ < |b| < 60^\circ$. Even within this sector, the ISM is not homogeneous and has a number of phases, each with a different \HI density and presenting a different environment for H-\Hbar oscillations.

In App.~\ref{app:ism}, we describe these phases and explain why we expect most of the oscillation-induced $\gamma$-rays to come from three of them, namely the cold neutral medium (CNM), warm neutral medium (WNM) and warm ionised medium (WIM). Here we present a short description of these three phases, as well as the nominal values we assume for their physical properties  \cite{Ferriere:2001rg, cox2005three, tielens2005physics, draine2010physics}. $T$ here represents the phase temperature, and $x$ the ionisation fraction.
\begin{itemize}
\item CNM: Comprises clumps of cold \HI clouds.\\
$n_{\mH} \simeq 50 \, \text{cm}^{-3}$, $T \simeq 80 \, \text{K}$, $x=0.001$.

\item WNM: Intercloud region containing warm diffuse \HI.\\
$n_{\mH} \simeq 0.5 \, \text{cm}^{-3}$, $T \simeq 8000 \, \text{K}$, $x=0.05$.

\item WIM: Intercloud region containing warm diffuse \HII.\\
$n_{\mHplus} \simeq 0.3 \, \text{cm}^{-3}$, $T \simeq 8000 \, \text{K}$, $x=0.9$.
\end{itemize}
The uncertainties in these nominal values, in particular the ionisation fraction, is a significant source of error in our analysis. Henceforth, most values that we present should only be interpreted as \emph{order-of-magnitude estimates}.


\subsection{Emissivities of the CNM, WNM and WIM}

We now want to determine the oscillation-induced emissivities of the three phases. To do so, we first need the values of $\epsilon'$, $\bar{\omega}_I$ and $\bar{\omega}_{\text{ann}}$ used in the emissivity formula Eq.~(\ref{eq:emissivity}). The values we present below incorporate a wide range of elastic targets as well as inelastic processes, using available data on scattering phase shifts, cross-sections and reaction rate constants \cite{schwartz1961electron, Armstead:1968zz, bhatia1971generalized, Morgan:1973zz, bhatia1974rigorous, kolos1975hydrogen, register1975algebraic, fon1978elastic, morgan1989atomic, mitroy1993close, krstic1999atomic,  sinha2004total, chakraborty2007cold} (more details can be found in App.~\ref{app:parameter}):

\begin{itemize}
  \item CNM:\\
  $\epsilon' \simeq (-1 \pm i) \times 10^{-7} \, \text{s}^{-1}$, mostly from elastic scattering with H.\\
  $\bar{\omega}_I \simeq \bar{\omega}_{\text{ann}} \simeq 6 \times 10^{-8} \, \text{s}^{-1}$, mostly from H-\Hbar annihilation.

  \item WNM:\\
  $\epsilon' \simeq (-5 \pm 5i) \times 10^{-9} \, \text{s}^{-1}$, mostly from elastic scattering with H.\\
  $\bar{\omega}_I \simeq \bar{\omega}_{\text{ann}} \simeq 8 \times 10^{-10} \, \text{s}^{-1}$, mostly from H-\Hbar annihilation.

  \item WIM:\\
  $\epsilon' \simeq (-2 - i) \times 10^{-8} \, \text{s}^{-1}$, mostly from elastic scattering with \eminus.\\
  $\bar{\omega}_I \simeq \bar{\omega}_{\text{ann}} \simeq 7 \times 10^{-10} \, \text{s}^{-1}$, mostly from \Hplus-\Hbar annihilation.
\end{itemize}
Our estimate for $\epsilon'$ are a few orders of magnitude larger than in \cite{Feinberg:1978sd}, where it was assumed that $2|\epsilon'| \simeq \bar{\omega}_I \simeq 10^{-10} \, \text{s}^{-1}$. This discrepancy is mainly due to contributions from elastic scattering that they have neglected. Hence, their assumption that H-\Hbar oscillations are mainly suppressed by \Hbar annihilation is not justified.

With these values, we can finally obtain the following oscillation-induced $\gamma$-ray emissivities per H atom.
\begin{itemize}
\item CNM: $\epsilon_{\gamma} \simeq 2g_\gamma|\delta|^2 \times 10^5 \, \text{s} \, \text{srad}^{-1}$.
\item WNM: $\epsilon_{\gamma} \simeq 4g_\gamma|\delta|^2 \times 10^6 \, \text{s} \, \text{srad}^{-1}$.
\item WIM: $\epsilon_{\gamma} \simeq g_\gamma|\delta|^2 \times 10^6 \, \text{s} \, \text{srad}^{-1}$.
\end{itemize}

Since the $\gamma$-ray data in \cite{Abdo:2009ka} starts at $100 \, \text{MeV}$, using the experimental and simulation results in \cite{Backenstoss:1983gu}, we estimate the average number of photons from \Hbar annihilation above this threshold to be $g_\gamma \simeq 2.7$.


\section{Deriving bound on $|\delta|$ using \textit{Fermi} LAT data} \label{sec:bounds}

In this section, we explain how we derive a bound on the oscillation parameter $|\delta|$ using \textit{Fermi} LAT data. The main idea is to compare the results of $\gamma$ ray measurements with predictions from astrophysical models. The difference between them can then be used to constrain additional oscillation-induced emissivity and hence $|\delta|$.

More specifically, one can perform a linear regression of the observed $\gamma$ ray intensity against the \HI column density. The slope corresponds to the emissivity per H atom, and the offset (intercept) a spatially homogeneous source of emissivity. The observed slope can be compared with independent astrophysical predictions to constrain $|\delta|$, and this was indeed what was done in \cite{Feinberg:1978sd}. However, we argue that the oscillation-induced emissivity should really show up in the offset rather than the slope, which lacks an independent prediction. Therefore, the whole measured offset is used to constrain $|\delta|$. We explain these points in more details below.

\subsection{Review of relevant $\gamma$-ray data}

In this section we review the analysis and results in \cite{Abdo:2009ka}. One of their goals was to determine the \HI $\gamma$-ray emissivity, and compare it with predictions based on CR interaction with matter. The authors used \textit{Fermi} LAT $\gamma$-ray data from the sector we previously described, in the energy range $100 \, \text{MeV}-9.05 \, \text{GeV}$. This sector is known to be free of large molecular clouds. In this region, \HII column-density is relatively smooth and is in the range $(1-2) \times 10^{20} \, \text{cm}^{-2}$, while \HI distribution is more clumpy with a column density in the range $(1-18) \times 10^{20} \, \text{cm}^{-2}$.

Known background such as point sources and inverse Compton scattering of soft photons with CR electrons were subtracted, leaving only data that are expected to come from CR interaction with matter as well as an isotropic extragalactic diffuse background. By comparing the post-subtraction $\gamma$-ray intensity map (Fig.~1 of \cite{Abdo:2009ka}) with a \HI column density map derived from $21\, \text{cm}$ radio surveys (Fig.~3 of \cite{Abdo:2009ka}), the authors found a linear relationship between the $\gamma$-ray intensity $I_\gamma$ and the \HI column density $N(\textsc{Hi})$ for each energy bin, which we index by $i$ (Fig.~4 of \cite{Abdo:2009ka})
\beq
I_{\gamma,i} \approx S_i \cdot N(\textsc{Hi}) + O_i
\eeq
where the slope $S_i$ represents the \HI emissivity per atom, and the offset $O_i$ the contributions from residual particles and the extragalactic background. The authors found good agreement between the slope-derived \HI emissivity and the predictions based on CR interaction with matter.
Summing the results in Tab.~1 of \cite{Abdo:2009ka} over the bins in the energy range $100 - 1130 \, \text{MeV}$ (relevant for \Hbar annihilation), we find that the \HI emissivity given by the combined slopes is 
\begin{equation}
S = 1.5 \times 10^{-26} \text{ photons } \, \text{s}^{-1} \, \text{sr}^{-1} \text{~per H atom}.
\end{equation}
and the combined offset is
\begin{equation} \label{eq:val-of-O}
O= 1.4 \times 10^{-5} \text{ photons } \text{cm}^{-2} \, \text{s}^{-1} \, \text{sr}^{-1}.
\end{equation}

\subsection{Bounds on $|\delta|$} \label{sec:scenario}

Let us now consider what happens if there are extra oscillation-induced $\gamma$-rays on top of the known sources. Distribution-wise, both the WIM and WNM have relatively low volume densities and large volume filling factors, so their contributions to the \HI column density should be relatively uniform over the column density map. In contrast, the CNM is clumpy with much higher density and smaller filling factor, so the small regions in the map with high column densities probably correspond to lines of sight which pass through the CNM. In other words, lines of sight with more H from the CNM provide the high leverage points that determine the slope in the linear regression of emissivity against column density. On the other hand, as we have seen, the extra emissivity per H atom varies among the three phases of ISM, with the WNM and WIM values being one order of magnitude higher than the CNM. Together, this suggests that the extra $\gamma$-ray intensity is more likely to show up in Fig.~4 of \cite{Abdo:2009ka} as a contribution to the offset rather than the slope.

\begin{figure}[h]
\centering
\includegraphics[width=0.6\linewidth]{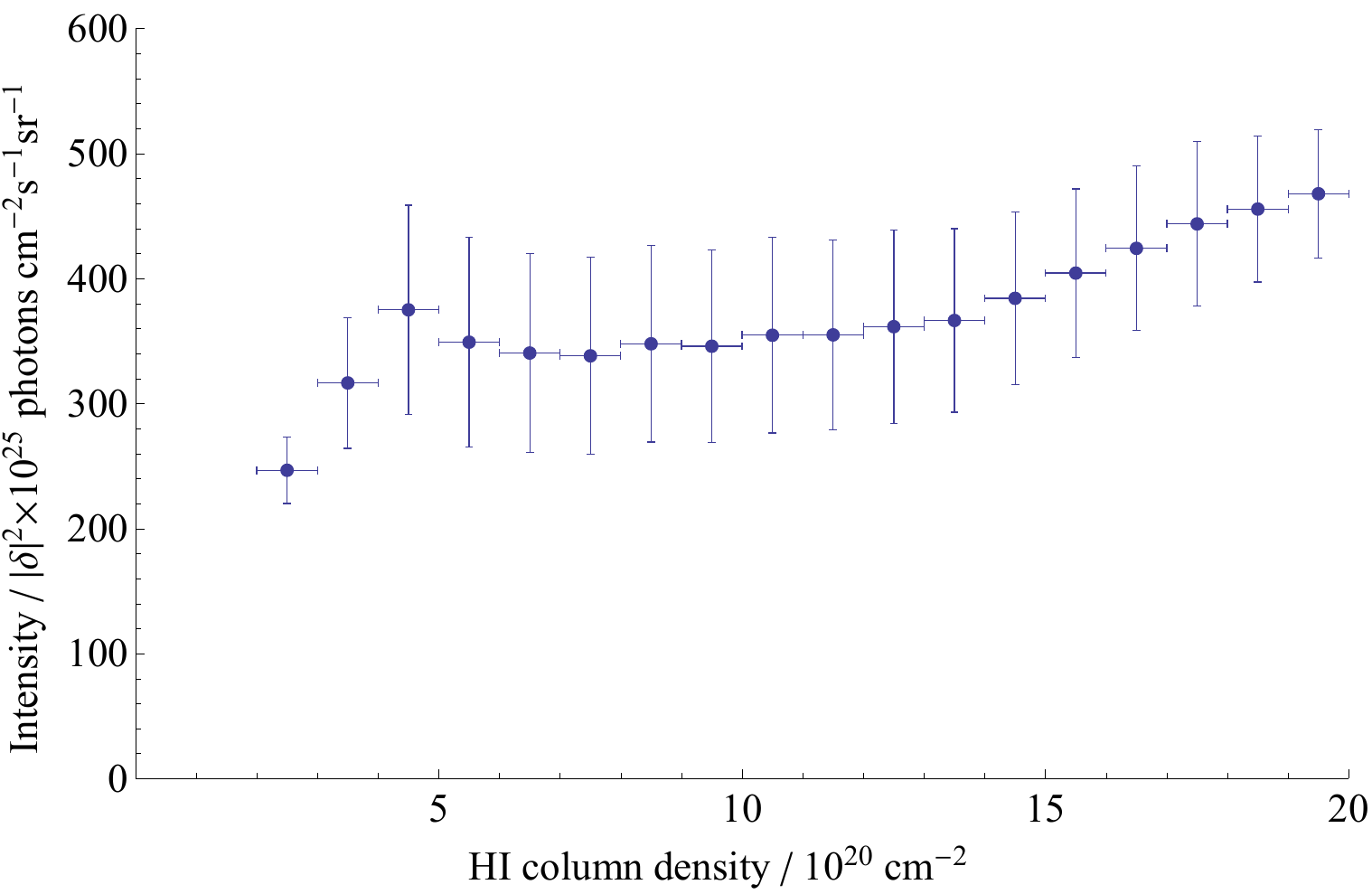}
\caption{Results of a simple calculation showing how the oscillation-induced $\gamma$-ray intensity varies with the \HI column density.} \label{fig:scenario}
\end{figure}

We perform a simple calculation to show that this is indeed the case. The WNM and WIM are assumed to be layers parallel to the galactic disk. Therefore, their contributions to the \HI column density are constant, except for a $\tfrac{1}{\sin|b|}$ latitudinal variation since a more ``glancing'' line of sight travels a longer distance through the layer. Using Eq.~(\ref{eq:cnm-wnm}) and (\ref{eq:wim}) and the nominal ionisation fraction, this corresponds to a contribution of $\tfrac{1.7}{\sin|b|} \times 10^{20} \, \text{cm}^{-2}$ from the WNM and $\tfrac{0.08}{\sin|b|} \times 10^{20} \, \text{cm}^{-2}$ from the WIM. On top of that, the CNM is assumed to add a random contribution that ranges from 0 to $\tfrac{10}{\sin|b|} \times  10^{20} \, \text{cm}^{-2}$. For each line of sight within the latitudinal range of interest, we calculate the total \HI column density and oscillation-induced $\gamma$-ray intensity, repeated many times over different random CNM contributions. Fig.~\ref{fig:scenario} shows a plot of intensity against column density, with the horizontal errorbars indicating the bin intervals, and the vertical errorbars the intensity range of the corresponding bins. The plot is mostly horizontal, indicating that the extra intensity is indeed more likely to show up in the offset, with a contribution of roughly
\beq O_{\text{osc.}} \simeq 4|\delta|^2 \times 10^{27} \text{ photons } \text{cm}^{-2} \, \text{s}^{-1} \, \text{sr}^{-1}. \label{eq:val-of-O-ind}\eeq

To obtain a bound on $|\delta|$, we identify this extra offset with the entire experimental offset value, which we found earlier to be around $1.4 \times 10^{-5} \text{ photons } \text{cm}^{-2} \, \text{s}^{-1} \, \text{sr}^{-1}$. In principle, we could have performed further background subtraction from this experimental value before making the identification. Possible background includes CR interaction with smoothly-distributed residual particles such as \HII, incomplete earlier subtraction of inverse Compton scattering due to model uncertainties, as well as extragalactic sources. However, these contributions are either not well-quantified, or turn out to be small compared to the experimental value, so the subtraction is unlikely to have made a big difference. Comparing $O$ and $O_{\text{osc.}}$ from Eqs.~(\ref{eq:val-of-O}) and (\ref{eq:val-of-O-ind}), we find that
\begin{equation}
|\delta| \lesssim 6 \times 10^{-17} \, \text{s}^{-1}.
\end{equation}
This is about one order of magnitude weaker than the bound derived in \cite{Feinberg:1978sd}. In other words, the earlier bound may have been too stringent. We also note that \cite{Feinberg:1978sd} used the slope (from older $\gamma$-ray data \cite{Fichtel:1977}) instead of the offset to derive the bound, so it did not account for the most likely scenario in which the CNM is mainly responsible for the variation in \HI column density from which the slope is derived, whereas the WNM dominates the extra oscillation-induced intensity.


\section{Discussion and conclusions} \label{sec:discussion}

The bounds we have derived on $|\delta|$ can be translated to a bound on four-fermion contact operators involving protons and electrons. For instance, \cite{Feinberg:1978sd} considered the operator
\begin{equation}
\mathcal{O}_1 = \frac{1}{\Lambda^2} [\bar{p}^c \gamma_\mu (1+\gamma_5) e][\bar{p}^c \gamma^\mu (1+\gamma_5) e] + \text{h.c.},
\end{equation}
and found that $\delta$ is related to $\Lambda$ via
\begin{equation}
\delta = \frac{16}{\Lambda^2 \pi a^3}, \label{eq:delta-from-cutoff}
\end{equation}
where $a$ is the Bohr radius.

On the other hand, $ppee$ operators can also be constrained by other processes such as $pp\to ee$. For instance, results from Super-Kamiokande can be used to set an upper bound on the proton annihilation rate in oxygen nuclei. For a benchmark operator
\begin{equation}
\mathcal{O}_2 = \frac{1}{\Lambda^2} (i\bar{p}^c \gamma_5 p)(i \bar{e}^c \gamma_5 e) + \text{h.c.},
\end{equation}
this translates to a bound of $\Lambda > 7 \times 10^{14} \,\text{GeV}$ \cite{Bramante:2014uda}. If we now assume that the same cutoff scale can be used in Eq.~(\ref{eq:delta-from-cutoff}) to estimate a bound on $|\delta|$, we find that
\begin{equation}
|\delta| \lesssim 10^{-21} \, \text{s}^{-1},
\end{equation}
which is actually four orders of magnitude more stringent than the bound that we have obtained from $\gamma$-ray observations.

It is unlikely that choosing a different region for $\gamma$-ray observations can give an improved bound on $|\delta|$ that is just as competitive, so it is worth speculating whether a terrestrial laboratory-based oscillation experiment might do better. For instance, if a falling H atom oscillates partially into an \Hbar, the experiment can attempt to detect $\gamma$-rays from annihilation when this atom comes into contact with a solid surface. Compared to measurements based on the ISM, the advantages are that annihilation no longer relies on chance encounters with other atoms, and that the $\gamma$-rays background can potentially be controlled. If there are $N$ H atoms each with a characteristic flight time $t$ before reaching a solid surface, then the absence of $\gamma$-rays indicate a crude bound of $(|\delta| t)^2 \lesssim \tfrac{1}{N}$. Unfortunately, even obtaining a bound close to that from the ISM is unlikely to be feasible. For instance, a bound of $|\delta| \lesssim 10^{-16} \, \text{s}^{-1}$, assuming a flight time of $t = 1 \, \text{s}$, will require about $10^8 \, \text{mol}$ of atomic hydrogen, a very large number. In addition, there are practical concerns about how rarefied the H atoms should be so that they do not start to interact, and the cryogenics required so that thermal motion does not substantially reduce the flight time.

To conclude, we have updated the bounds on H-\Hbar oscillations based on oscillation-induced $\gamma$-ray emission in the ISM. Suppression from elastic collisions turn out to be more significant than assumed in previous work, and using a multi-phase ISM model as well as updated parameter values and $\gamma$-ray data, we show that the upper bound on $|\delta|$ is about $6 \times 10^{-17} \, \text{s}^{-1}$, one order of magnitude weaker than previously thought.

\begin{acknowledgments}
We thank Lorenzo Calibbi, Shmuel Nussinov and Chelsea Sharon for helpful discussions. Work of YG is supported in part by the NSF grant PHY1316222. Work of SR is supported by the Department of Science and Technology, Government of
India through INSPIRE Faculty Fellowship (Grant no. IFA12-PH-41).
\end{acknowledgments}

\newpage

\appendix

\section{More details about the H-\Hbar oscillation model}

\subsection{Elastic scattering} \label{app:formalism}

The model we used in this work was originally derived in \cite{Feinberg:1961zza} somewhat heuristically based on the notion of a classical sum over different ``histories'', where in each infinitesimal time interval $\delta t$, the atom may undergo either elastic scattering or quantum time evolution. We have been able to re-derive the model on a more rigourous basis as follows.

The atom is originally described by a density matrix in the product space $\mathcal{H}_A \otimes \mathcal{H}_B$, where $\mathcal{H}_A$ is associated with the atom's identity, and $\mathcal{H}_B$ with momentum degrees of freedom (for simplicity we neglect atomic level and spin; including them simply increases the number of Wigner functions). We then extend the impurity-scattering formalism described in \cite{rammer2004quantum} to derive quantum kinetic equations for the $2 \times 2$ Wigner functions. By making a number of assumptions before and after integrating over momentum space (equivalent to tracing out $\mathcal{H}_B$), we finally obtain the same kinetic equation for the reduced $2\times 2$ density matrix $\rho(t)$ as \cite{Feinberg:1961zza}.

We now examine the various assumptions made in this derivation.
\begin{itemize}
\item The derivation of the Wigner function kinetic equations assumed that the mean free path be much larger than the de Broglie wavelength, and that quantum degeneracy as well as two-body correlation between atom and target can be ignored. These are probably reasonable assumptions for an atom in the ISM.

\item In further reducing these kinetic equations to the one for $\rho(t)$, two further assumptions are made. First, we take the classical limit of the scattering terms, which requires that memory effects be neglected, again a reasonable assumption given that the momentum relaxation time of an atom is much shorter than our timescale of interest (the quasi-steady rate of change). Second, as mentioned in Sec.~\ref{sec:model-description}, in order to close the kinetic equation for $\rho(t)$, we assume that moments in momentum and other degrees of freedom can be replaced by products of $\rho(t)$ with the relevant expectation values. While some errors are introduced in doing so, they are not expected to be very significant.

\item The impurity-scattering formalism assumes that the targets are immobile, certainly not true for real targets in the ISM. Nonetheless, this can be addressed by replacing $v$ and $p$, not by the r.m.s.~values in the lab frame, but rather the r.m.s.~values evaluated in the two-particle centre-of-mass frame comprising the atom and a target particle (hence this also involves averaging over the target velocity distribution). Only $E$ should still be the lab frame value.

\item Finally, the impurity-scattering formalism assumes that the atom and target are distinguishable particles. This is clearly violated if we consider scattering with other H atoms. Both $f(\theta)$ and $f(\pi - \theta)$ will then contribute to the same H-H scattering process, and one must also be careful not to double-count the phase space. This is probably the biggest source of error (possibly up to a factor of 2) in the model, at least for the CNM and WNM. However, there is not much point in trying to derive a more accurate treatment due to the lack of accurate scattering data.
\end{itemize}

\subsection{Inelastic processes} \label{app:inelastic}

In Sec.~\ref{sec:inelastic}, we only considered inelastic processes where the H/\Hbar atom ``disappears'', e.g.~\Htwo formation or \Hbar annihilation. These processes cause the state to leave the Hilbert space $\mathcal{H}_A$ and can hence be represented by imaginary diagonal contributions to the effective Hamiltonian. However, there are other processes where the atom does not disappear but are nonetheless inelastic. We now explain why they can be neglected.

First, we consider processes like $\mH/\mHbar(1S) + X \to \mH/\mHbar(1S) + Y$, where the H/\Hbar atom remains in the $1S$ state but the target is collisionally excited/ionised/dissociated. As far as the H/\Hbar atom is concerned, these processes are not very different from elastic scattering, and so enters the model in a similar manner (except without a forward scattering contribution). However, we expect them to be less important than elastic scattering off the same target $X$ since the rates are usually Boltzmann-suppressed in comparison, even in the warm phases.

Next, we consider collisional and photo-excitations of H/\Hbar to $n \ge 2$ atomic states. These processes (together with collisional and radiative decays) are responsible for maintaining the quasi-steady distribution of atomic levels. However, if the transition amplitudes for H and \Hbar are different, then one also needs to examine how they might directly affect the oscillations. Collisional excitations can again be neglected since they are Boltzmann-suppressed compared to elastic scattering. For photo-excitations, the electric dipole transition amplitudes for H and \Hbar do indeed differ by a sign; however, there is hardly any time for the $\mathcal{H}_A$ part of the state to evolve (except by an overall phase) before the atom undergoes radiative decay that undoes the sign change. Therefore, the net direct effects are also unimportant.

The arguments above do not apply to $1S$ hyperfine transitions. In particular, collisional excitations to the higher-energy hyperfine state are not Boltzmann-suppressed. However, since these processes involve electron spin flips, they are either magnetic in nature and hence have smaller cross-sections, or rely on electron exchange (e.g.~when the target is \eminus or other H atoms) and  hence already included in conventional elastic scattering data. Photo-excitations can also occur via dipole transition to $nP$ states followed by decays to the higher $1S$ hyperfine state, but as explained above the net direct effects are unimportant due to sign cancellation.

\section{Phases of the ISM} \label{app:ism}

The ISM comprises a number of phases that accounts for most of its mass and volume. Parameter values are taken from \cite{Ferriere:2001rg, cox2005three, tielens2005physics, draine2010physics}.
\begin{itemize}

\item Neutral atomic gases: There are two phases that contain predominantly \HI. The CNM comprises \HI clouds typically of size $\mathcal{O}(10) \, \text{pc}$, number density $20 - 50 \, \text{cm}^{-3}$, temperature $50 - 100 \, \text{K}$ and volume filling factor $\mathcal{O}(0.01)$. The WNM comprises diffuse intercloud \HI, typically with a lower number density $0.2 - 0.6 \, \text{cm}^{-3}$, and higher temperature $5000 - 10000 \, \text{K}$ and filling factor $0.3-0.4$. Locally, a simple model for the vertical \HI distribution (filling factor incorporated) is given by
\begin{equation}
n_{\mH}(z) / \text{cm}^{-3} = 0.40 e^{-\left(\frac{z}{127 \, \text{pc}}\right)^2} + 0.10 e^{-\left(\frac{z}{318 \, \text{pc}}\right)^2} + 0.063 e^{-\frac{|z|}{403 \, \text{pc}}},
\label{eq:cnm-wnm}
\end{equation}
where the first term corresponds to the CNM, and the second and third terms the WNM.

\item Warm ionised gases: Radiation from O and B stars cause almost-complete ionisation of nearby clouds, so most of the hydrogen are in the ionised form \HII. These \HII regions, typically of size $\mathcal{O}(1) \, \text{pc}$, are generally very dense and hot, with number densities up to $\mathcal{O}(10^5) \, \text{cm}^{-3}$, temperatures $8000 - 10000 \, \text{K}$, and negligibly small filling factors. Besides these dense regions, there also exists a diffuse warm ionised phase called the WIM. This phase has comparable temperature, but much lower number density $\sim 0.1 - 0.5 \, \text{cm}^{-3}$, and much higher filling factor $0.05 - 0.25$. A simple ``two-disk'' model for the vertical \HII distribution is given by
\begin{equation}
n_{e}(z) / \text{cm}^{-3} = 0.015 e^{-\frac{|z|}{70 \, \text{pc}}} + 0.025 e^{-\frac{|z|}{900 \, \text{pc}}},
\label{eq:wim}
\end{equation}
where the first term represents the collection of localised \HII regions as a ``thin-disk'', and the second term the WIM as a ``thick disk''.

\item Coronal gases: Besides the WIM, there is another diffuse ionised phase referred to as coronal gases, because the temperature and ionisation state are believed to be similar to that of the solar corona. This phase is much hotter and rarefied, with temperature $\mathcal{O}(10^5 - 10^6) \, \text{K}$, number density $0.003 - 0.007 \, \text{cm}^{-3}$, and filling factor $0.2 - 0.5$. The vertical profile depends on the measurements used (e.g.~choice of spectral lines) but usually fits a large scale height of 3 kpc (assuming exponential distribution) or above.

\item Molecular clouds: These comprise gravitationally-bound clouds, typically of size $\mathcal{O}(10) \, \text{pc}$ with \Htwo as the dominant species. They are typically very cold and dense, with temperature $10 - 20 \, \text{K}$, number density up to $\mathcal{O}(10^6) \, \text{cm}^{-3}$, and negligible filling factor. Vertically, they tend to be concentrated near the galactic disk, with a Gaussian scale height around $70 - 80 \, \text{pc}$.

\end{itemize}

While the main constituents in these phases are H, \Htwo, \Hplus and \eminus, also present are other gaseous elements and dust.
\begin{itemize}
\item Other gaseous elements: From photospheric and meteoritic measurements, the cosmic composition in terms of number density are as follows: He $10\%$, C $0.03\%$, O $0.05\%$, and all other species individually each below $0.01\%$ (combined $\sim 0.03\%$). There is also evidence that a significant fraction of these elements might have been locked up in dust and hence depleted in the gaseous form.

\item Dust: Dust grains are generally well-mixed with the gases in the ISM, with a dust-to-gas mass ratio believed to be around $\mathcal{O}(0.01)$. The dust grains are primarily composed of heavier elements like C, N, O, Mg, Si and Fe, with a typical specific density of $3 \, \text{g} \, \text{cm}^{-3}$. A popular model for the grain-size distribution (based on the extinction curve) is the Mathis-Rumpl-Nordsieck model. In the model, the dust grains are assumed to graphite and silicates, and the distribution given by
\begin{equation}
n_i(a) da = A_i n_{\mH} a^{-3.5} da,
\end{equation}
where $a$ is the grain size, and $A_i$ is $7.8 \times 10^{-26}$ and $6.9 \times 10^{-26} \, \text{cm}^{2.5}$ for silicates and graphite respectively. This relation holds over the range $50 \, \text{\AA} < a < 2500 \, \text{\AA}$. Besides large dust grains, it is also believed that there exists a population of large polycyclic aromatic hydrocarbon molecules, with an relative abundance of $\mathcal{O}(10^{-5})\%$.
\end{itemize}

Having described the phases of the ISM, we now argue that we only need to consider oscillation-induced $\gamma$-ray contributions from the CNM, WNM and WIM. For instance, consider the dense molecular clouds. Looking at Eq.~(\ref{eq:emissivity}), since most contributions to $\epsilon'$, $\bar{\omega}_I$ and $\bar{\omega}_{\text{ann}}$ scale roughly with the gas density, this means that the emissivity per H atom is much smaller than in the more rarefied phases. While the gas column density may be very high along lines of sight passing through the clouds, only a tiny fraction of the gas is \HI, so this is unlikely to compensate for the lower emissivity per H atom. In addition, \cite{Abdo:2009ka} specifically mentions that large molecular clouds are known to be absent in the sector of interest. Similar types of arguments can also be made for the dense \HII regions and the coronal gases to explain why they can be neglected.

\section{Parameter values} \label{app:parameter}

We present here a summary of the contributions from both elastic and inelastic processes to the parameters $\epsilon'$, $\bar{\omega}_I$ and $\bar{\omega}_{\text{ann}}$. Properties of the three phases are assumed to follow the nominal values given in Sec.~\ref{sec:ism}.

\subsection{Elastic scattering}

From Eq.~(\ref{eq:epsilonprime}), recall that the contribution of elastic scattering to $\epsilon'$ from target species $i$ is given by
\begin{equation}
\Delta \epsilon' =  n_i v_i \left\lbrace - \int d\Omega \tfrac{|f_{i,p_i} - \bar{f}_{i,p_i}|^2}{2} + i\left[ \tfrac{2\pi\text{Re}[f_{i,p_i}(0)-\bar{f}_{i,p_i}(0)]}{p_i} + \int d\Omega \text{Im}(\bar{f}_{i,p_i}^*f_{i,p_i}) \right]\right\rbrace.
\end{equation}
We now calculate this contribution for different target species.

\subsubsection{\eminus as targets}

It is useful to begin with elastic (H/\Hbar)-\eminus scattering for the WNM and WIM (we neglect the CNM due to its extremely low ionisation fraction). First, amplitude data are available for both H and \Hbar. Second, \eminus may potentially be the dominant target species, since the much lower reduced mass (around $m_{\text{e}}$) implies a higher speed of approach $v$ and smaller centre-of-mass momentum $p$, hence boosting $\Delta\epsilon'$.

For H-\eminus partial wave phase shifts, we use \cite{schwartz1961electron, Armstead:1968zz,fon1978elastic}, while for \Hbar-\eminus phase shifts, we use \cite{bhatia1971generalized, bhatia1974rigorous, register1975algebraic, mitroy1993close}. At the warm phase temperature (about $1 \, \text{eV}$), we find that
\begin{equation}
\begin{aligned}
\tfrac{1}{4} \int d\Omega \tfrac{|f_s - \bar{f}|^2}{2} + \tfrac{3}{4} \int d\Omega \tfrac{|f_t - \bar{f}|^2}{2} &\simeq 13 \, \text{\AA}^2,\\
\tfrac{1}{4} \tfrac{2\pi\text{Re}[f_s(0)-\bar{f}(0)] }{p} + \tfrac{3}{4} \tfrac{2\pi\text{Re}[f_t(0)-\bar{f}(0)]}{p} &\simeq -11 \, \text{\AA}^2,\\
\tfrac{1}{4} \int d\Omega \text{Im}(f_s \bar{f}) + \tfrac{3}{4} \int d\Omega \text{Im}(f_t \bar{f}) &\simeq 3.8 \, \text{\AA}^2,
\end{aligned}
\end{equation}
where $f_s$ and $f_t$ and are the electronic singlet and triplet H-\eminus amplitudes. To check that the first value makes sense, we note that the elastic H-\eminus singlet and triplet cross-sections ($39 \, \text{\AA}^2$ and $15 \, \text{\AA}^2$) are much larger than the \Hbar-\eminus cross-section ($1.6 \, \text{\AA}^2$). This suggests that $f_s, f_t \gg \bar{f}$, in which case the first value should be approximately half the spin-averaged H-\eminus cross-section. This gives a reasonably close value of $11\, \text{\AA}^2$.

For an r.m.s.~speed of approach $v  = \sqrt{\tfrac{3kT}{m_{\text{e}}}} \simeq 6 \times 10^7 \, \text{cm} \, \text{s}^{-1}$, we obtain
\begin{itemize}
\item WNM: $\Delta \epsilon' \simeq (-2 - i) \times 10^{-9} \, \text{s}^{-1}$.
\item WIM: $\Delta \epsilon' \simeq (-2 - i) \times 10^{-8} \, \text{s}^{-1}$.
\end{itemize}

\subsubsection{\Hplus as targets}

Next, we consider elastic (H/\Hbar)-\Hplus scattering, again for the WNM and WIM. Here, a number of issues arise. First, a much larger number of partial waves are required to accurately reconstruct the scattering amplitudes, since the centre-of-mass momentum $p$ is now much higher. For H-\Hplus scattering, while phase shifts for nearly 200 partial waves are available \cite{hunter1980scattering}, we found that they are nonetheless insufficient for the forward scattering amplitude\footnote{Recall that for partial wave amplitudes $a_l$, $f(0)$ involves a summation of $(2l+1)a_l$ as opposed to $(2l+1)|a_l|^2$ for the total cross-section, hence implying a slower convergence.}. Second, we have not been able to find scattering data for \Hplus-\Hbar scattering. Therefore, unlike the previous case, here an accurate calculation is not possible. The approach we adopt is as follows.

\cite{morgan1989atomic} claims that the elastic H-$\bar{\text{p}}$ (charge-conjugate of \Hbar-\Hplus) cross-section is comparable to the re-arrangement cross-section ($11 \, \text{\AA}^2$ from \cite{Morgan:1973zz}). Should this indeed be the case, this implies that the elastic \Hbar-\Hplus cross-section is much smaller than that of H-\Hplus ($160 \, \text{\AA}^2$ from \cite{krstic1999atomic} after nuclear-spin averaging). If we then assume that $\bar{f} \ll f$, we can drop $\bar{f}$ in the expression for $\Delta \epsilon'$, giving
\begin{equation}
\Delta \epsilon' \simeq  n v \left\lbrace - \int d\Omega \tfrac{|f|^2}{2} + i \tfrac{2\pi\text{Re}[f(0)]}{p}\right\rbrace,
\end{equation}
so only H-\Hplus data is required. The first term requires the nuclear-spin averaged cross-section, and the second term the averaged forward scattering amplitude.

Instead of the phase shifts from \cite{hunter1980scattering}, we mostly rely on the averaged differential and total cross-sections from \cite{krstic1999atomic}, since the latter is more recent and includes a larger number of partial waves (more than 500). To extract the averaged $\text{Re}[f(0)]$, we first note that the nuclear singlet and triplet amplitudes are given by $f_{s,t}(\theta) = f_d(\theta) \pm f_e(\pi - \theta)$, where $f_d$ and $f_e$ are the ``direct'' and ``charge exchange'' amplitudes had the nuclei been distinguishable \cite{krstic1999atomic}. At energies $\gtrsim 1\, \text{eV}$, both $f_d(\theta)$ and $f_e(\theta)$ become so forward-distributed that $f_s(0) \simeq f_t(0) \simeq f_d(0)$, while the overlap between $f_d(\theta)$ and $f_e(\pi - \theta)$ become so small that the singlet and triplet total cross-sections become identical. We then use the optical theorem to estimate $\text{Im}[f_d(0)]$ from the spin-averaged cross-section, which in turn can be used to estimate $|\text{Re}[f_d(0)]|$ from the spin-averaged differential cross-section at $\theta\simeq 0$. We only use the phase shifts from \cite{hunter1980scattering} to fix the sign of $\text{Re}[f_d(0)]$ and to check the validity of the assumptions above. We find that
\begin{align*}
\int d\Omega \tfrac{|f|^2}{2} &\simeq 81 \, \text{\AA}^2,\\
\tfrac{2\pi\text{Re}[f(0)]}{p} &\simeq 74 \, \text{\AA}^2,
\end{align*}
from which we obtain
\begin{itemize}
\item WNM: $\Delta \epsilon' \simeq (-4 + 4i) \times 10^{-10} \, \text{s}^{-1}$.
\item WIM: $\Delta \epsilon' \simeq (-5 + 4i) \times 10^{-9} \, \text{s}^{-1}$.
\end{itemize}
These $\Delta \epsilon'$ values are smaller than that of (H/\Hbar)-\eminus scattering, mostly due to the much smaller speed of approach $v$.

\subsubsection{H as targets}

Finally, we consider elastic (H/\Hbar)-H scattering for the CNM and WNM (we neglect the WIM due to its high ionisation fraction). We have not been able to find amplitude-level data, and even differential cross-section data is only limited to the WNM. Therefore, we will only perform a crude estimate of $\Delta \epsilon'$ using total cross-section data. We use \cite{krstic1999atomic} and \cite{chakraborty2007cold} for H-H and \cite{sinha2004total} for H-\Hbar cross-sections. Actually \cite{sinha2004total} only covers up to $0.27 \,\text{eV}$, a few times lower than the WNM temperature. However, since the cross-section appears relatively constant near $0.27 \, \text{eV}$, the cross-section should not differ significantly between $0.27 \, \text{eV}$ and $1\, \text{eV}$.

For H-H scattering, the CNM electronic singlet and triplet cross-sections are around $130 \, \text{\AA}^2$ and $60 \, \text{\AA}^2$, and the WNM spin-averaged cross-section $50 \, \text{\AA}^2$. For H-\Hbar scattering,  the CNM cross-section is $90 \, \text{\AA}^2$, and the WNM $60 \, \text{\AA}^2$. Based on these cross-sections, we now assume that $-\text{Re}(\Delta \epsilon') \simeq |\text{Im}(\Delta\epsilon')| \simeq n v (100 \, \text{\AA}^2)$ for the CNM, and $ n v (50 \, \text{\AA}^2)$ for the WNM. We then obtain
\begin{itemize}
\item CNM: $\Delta \epsilon' \simeq (-1 \pm i) \times 10^{-7} \, \text{s}^{-1}$.
\item WNM: $\Delta \epsilon' \simeq (-5 \pm 5i) \times 10^{-9} \, \text{s}^{-1}$.
\end{itemize}

\subsubsection{Other targets}

While other neutral targets such as He and \Htwo may offer slightly larger cross-sections than H, nonetheless their much lower abundances mean that their contributions to $\epsilon'$ can be ignored. The same can be said for other charged targets compared to \Hplus or \eminus.

\subsection{Inelastic processes}

For inelastic processes, we consider \Hbar annihilation, ionisation of H/\Hbar, as well as chemical reactions involving H. Keep in mind that $\omega_I$ only enters Eq.~(\ref{eq:emissivity}) as $\omega_I + \bar{\omega}_I$, so even the dominant contribution to $\omega_I$ can be ignored if it turns out to be much smaller than $\bar{\omega}_I$.

\subsubsection{\Hbar annihilation with H}

We use the semi-classical calculations of the rearrangement cross-section from \cite{kolos1975hydrogen}. Note that while there are fully-quantum calculations of the annihilation cross-section that include both rearrangement and annihilation-in-flight \cite{voronin1998antiproton, jonsell2001stability, armour2002calculation, armour2005inclusion}, they only include the $s$-wave component and hence give values that are much smaller. We now discuss each phase in turn.

\begin{itemize}
\item CNM: The cross-section is $\sigma \simeq 60 \, \text{\AA}^2$, corresponding to a rate coefficient of $\langle \sigma v \rangle \simeq 10^{-9} \, \text{cm}^3 \, \text{s}^{-1}$. The contribution to $\bar{\omega}_I$ is given by $n_{\mH}\langle \sigma v\rangle \simeq 6 \times 10^{-8} \, \text{s}^{-1}$.

\item WNM: The cross-section is $\sigma \simeq 8 \, \text{\AA}^2$, corresponding to a rate coefficient of $\langle \sigma v \rangle \simeq 2 \times 10^{-9} \, \text{cm}^3 \, \text{s}^{-1}$. The contribution to $\bar{\omega}_I$ is given by $n_{\mH}\langle \sigma v\rangle \simeq 8 \times 10^{-10} \, \text{s}^{-1}$.
\end{itemize}
We ignore this for the WIM due to the high ionisation fraction.

\subsubsection{\Hbar annihilation with \Hplus}

We again use semi-classical calculations from \cite{Morgan:1973zz}, since more updated cross-sections are either again for $s$-waves \cite{voronin1998antiproton}, or do not fully cover our energy range of interest \cite{sakimoto2001protonium, sakimoto2001full}. (In any case, we note that discrepancies between \cite{Morgan:1973zz} and \cite{sakimoto2001protonium, sakimoto2001full} where they do overlap are rather small.)

We ignore this for the CNM due to the extremely low ionisation fraction. For the WNM and WIM, we find a cross-section of $\sigma = 10 \, \text{\AA}^2$, corresponding to a rate coefficient of $\langle \sigma v \rangle \simeq 2 \times 10^{-9} \, \text{cm}^3 \, \text{s}^{-1}$. Hence we obtain the following results.
\begin{itemize}
\item WNM: The contribution to $\bar{\omega}_I$ is $n_{\mHplus}\langle \sigma v\rangle \simeq 6 \times 10^{-11} \, \text{s}^{-1}$.

\item WIM: The contribution to $\bar{\omega}_I$ is $n_{\mHplus}\langle \sigma v\rangle \simeq 7 \times 10^{-10} \, \text{s}^{-1}$.

\end{itemize}

\subsubsection{Other \Hbar annihilation processes}

One might expect \eminus-\Hbar annihilation to be important (especially in the WIM) since the relative speed $v$ is much higher. However, the annihilation cross-section turns out to be much smaller, due to the $6.8 \,\text{eV}$ energy threshold for re-arrangement, and that direct annihilation-in-flight in this case involves the electromagnetic interaction as opposed to the strong interaction \cite{armour2005inclusion}.

Finally, annihilation of \Hbar with any other neutral or charged species is expected to be less important than with H or \Hplus, due to their much lower abundances.

\subsubsection{Ionisation}

Ionisation in the \HI phases proceeds mainly via CR ionisation, at a rate per atom of order $10^{-16} \, \text{s}^{-1}$ \cite{tielens2005physics, draine2010physics}. For the WIM, photo-ionisation plays the more important role \cite{tielens2005physics}. A reasonable ionisation rate per atom in the WIM is $\mathcal{O}(10^{-13} - 10^{-12}) \, \text{s}^{-1}$, consistent with the degree of ionisation given typical recombination rates, as well as estimates of the ionisation parameter based on spectral measurements. Nonetheless, we see that in all three phases, the ionisation rates are much smaller than the contributions to $\bar{\omega}_I$ from \Hbar annihilation.

\subsubsection{Chemical reactions}

Many chemical reactions involve H and may contribute to $\omega_I$. However, all the rates are much smaller than $\bar{\omega}_I$, either because they involve species with very low abundances, or that they have very small rate coefficients. We discuss a number of examples here. The rate coefficients are taken from \cite{tielens2005physics}.

\begin{itemize}
\item Neutral reaction $\mH + \text{CH} \to \text{C} + \mHtwo$ has a rate coefficient $k = 1.2 \times 10^{-9} \left(\tfrac{T}{300 \, \text{K}}\right)^{0.5} e^{-\tfrac{2200 \, \text{K}}{T}}$. Even in the warm phases where the exponential suppression (from the activation barrier) becomes insignificant, the rate per H atom remains small due to the low abundance of CH.

\item \Htwo formation through $\mH + \text{H}^- \to \mHtwo + \meminus$ has a high rate coefficient $k = 1.3 \times 10^{-9} \, \text{cm}^3 \, \text{s}^{-1}$, but the \Hminus abundance is very low.

\item Radiative association $\mH + \meminus \to \mHminus + \gamma$ has a very low rate coefficient $k = 10^{-18} \tfrac{T}{1 \, \text{K}} \, \text{cm}^3 \, \text{s}^{-1}$.

\item Radiative association $\mH + \mH \to \mHtwo + \gamma$ has a very low rate coefficient $k \lesssim 10^{-23} \, \text{cm}^3 \, \text{s}^{-1}$.

\item Accretion of H on dust grain surface (an important catalytic reaction for \Htwo formation) occurs at a very low rate of roughly $10^{-17} \left(\tfrac{T}{10 \, \text{K}}\right)^{0.5} n_{\mH} \, \text{s}^{-1}$ per atom. (The $n_{\mH}$ dependence comes from the assumption of a constant dust-to-gas mass ratio.)

\end{itemize}

\bibliography{paper}
\bibliographystyle{apsrev4-1}

\end{document}